\documentclass[journal]{IEEEtran}

\usepackage{amsmath}
\usepackage{graphicx}
\usepackage{url}
\usepackage{hyperref}
\usepackage{cite}
\usepackage{amsfonts}
\usepackage{mathtools}
\usepackage{hhline}
\usepackage[caption=false, font=footnotesize]{subfig}
\usepackage{tabstackengine}
\usepackage{algpseudocode}
\usepackage{algorithm}
\algnewcommand\algorithmicforeach{\textbf{for each}}
\algdef{S}[FOR]{ForEach}[1]{\algorithmicforeach\ #1\ \algorithmicdo}

\usepackage[table]{xcolor}
\usepackage{commath,multirow}
\usepackage{tabularx} 
\usepackage{diagbox}

\usepackage{verbatim} 

\usepackage{adjustbox,cleveref}

\newcommand{\etal}{\textit{et al}. }
\newcommand{\ie}{{i}.{e}.}
\newcommand{\eg}{{e}.{g}.}

\makeatletter
\long\def\@makecaption#1#2{\ifx\@captype\@IEEEtablestring%
\footnotesize\begin{center}{\normalfont\footnotesize #1}\\
{\normalfont\footnotesize\scshape #2}\end{center}%
\@IEEEtablecaptionsepspace
\else
\@IEEEfigurecaptionsepspace
\setbox\@tempboxa\hbox{\normalfont\footnotesize {#1.}~~ #2}%
\ifdim \wd\@tempboxa >\hsize%
\setbox\@tempboxa\hbox{\normalfont\footnotesize {#1.}~~ }%
\parbox[t]{\hsize}{\normalfont\footnotesize \noindent\unhbox\@tempboxa#2}%
\else
\hbox to\hsize{\normalfont\footnotesize\hfil\box\@tempboxa\hfil}\fi\fi}
\makeatother

\begin{document}
%
\title{Capturing Video Frame Rate Variations via Entropic Differencing}

\author{Pavan C. Madhusudana, Neil Birkbeck, Yilin Wang,  Balu Adsumilli and Alan C. Bovik 
	\thanks{P. C. Madhusudana and A. C. Bovik are with the Department of Electrical and
Computer Engineering, University of Texas at Austin, Austin, TX, USA (e-mail:
pavancm@utexas.edu; bovik@ece.utexas.edu). Neil Birkbeck, Yilin Wang
and Balu Adsumilli are with Google Inc. (e-mail: birkbeck@google.com; yilin@google.com; badsumilli@google.com).}}


\maketitle

\begin{abstract}
High frame rate videos are increasingly getting popular in recent years, driven by the strong requirements of the entertainment and streaming industries to provide high quality of experiences to consumers. To achieve the best trade-offs between the bandwidth requirements and video quality in terms of frame rate adaptation, it is imperative to understand the effects of frame rate on video quality. In this direction, we devise a novel statistical entropic differencing method based on a Generalized Gaussian Distribution model expressed in the spatial and temporal band-pass domains, which measures the difference in quality between reference and distorted videos. The proposed design is highly generalizable and can be employed when the reference and distorted sequences have different frame rates. Our proposed model correlates very well with subjective scores in the recently proposed LIVE-YT-HFR database and achieves state of the art performance when compared with existing methodologies.
\end{abstract}


\section{Introduction}
\IEEEPARstart{A}{s} current media technology continues to emphasize ever higher quality regimes and to involve more immersive and engaging experiences for consumers, the need to extend current video parameter spaces along spatial and temporal resolutions, screen sizes and dynamic ranges has become a topic of extreme importance, especially in the media and streaming industry. Existing and emerging standards have increasingly focused on improving spatial resolution (4K/8K) \cite{ge2017toward}, High Dynamic Range (HDR) \cite{mai2010optimizing,kundu2017no}, and multiview formats \cite{smolic2007coding,de2013toward}. However there has been much less emphasis placed on increasing frame rates, and for a long time the frame rates associated with television, cinema and other video streaming applications have changed little - rarely above 60 frames per second (fps).

Various factors have limited increased adoptions of High Frame Rate (HFR) videos. Switching to HFR requires employing complex capture and display technologies which were not commonly available. Another possible reason for the limited popularity of HFR relates to limited knowledge about the perceptual benefits of employing HFR, which partly arises due to insufficient availability of HFR content. Recently, HFR has gathered significant interest among the research community, along with publication of databases such as the Waterloo HFR \cite{nasiri2015perceptual}, BVI-HFR \cite{mackin2018study} and LIVE-YT-HFR \cite{pavan2020liveythfr} datasets that exclusively target HFR contents. 


Perceptual Video Quality Assessment (VQA) is an integral component in numerous video applications such as digital cinema, video streaming applications (such as YouTube, Netflix, Hulu etc.) and social media (Facebook, Instagram etc). VQA models can be broadly classified into three main categories \cite{chikkerur2011objective}: Full-Reference (FR), Reduced-Reference (RR) and No-Reference (NR) models. FR VQA models require entire pristine undistorted stimuli along with degraded versions \cite{wang2004image,wang2003multiscale,sheikh2006image,zhang2011fsim,seshadrinathan2009motion,vu2011spatiotemporal,manasa2016optical}, while RR models operate with limited reference information \cite{wang2005reduced,li2009reduced,soundararajan2012rred,soundararajan2012video,bampis2017speed}. NR models operate without any knowledge of pristine stimuli \cite{mittal2012no,mittal2013making,saad2014blind,li2016spatiotemporal}. This work addresses the problem of quality evaluation when pristine and distorted sequences can possibly have different frame rates, thus our primarily focus will be on FR and RR VQA methods. 


There has been very limited work done on addressing VQA in the HFR domain. One of the first models was proposed by Nasiri \etal \cite{nasiri2017perceptual}, where they measured the amount of aliasing occurring in the temporal frequency spectrum, employing that as a measure of quality. In \cite{nasiri2018temporal} a motion smoothness measure was proposed for cross frame rate quality evaluation. Zhang \etal \cite{zhang2017frame} proposed a wavelet domain based Frame Rate Quality Metric (FRQM), where the differences between the wavelet coefficients of reference and temporally upsampled distorted sequences were used to predict quality. FRQM has a restriction that it cannot be employed when the reference and distorted videos have same frame rate, thus limiting its generalizability. 

In this paper, we propose a statistical VQA model that can capture distortions arising due to frame rate variations, and provide quality predictions that correlate well with human perception. This model is primarily motivated by temporal variations observed in the distributions of band-pass coefficients. We propose a novel entropic differencing method using Generalized Gaussian Distribution (GGD) model for both spatial and temporal band-pass responses, and show its effectiveness in capturing spatio-temporal artifacts. We evaluate our model on the LIVE-YT-HFR database and show that the predicted quality estimates have superior correlations against human judgments as compared to existing methods. Our proposed method is simplistic in nature, has very few hyperparameters to tune and does not require any computationally intensive training process. 

The rest of the paper is organized as follows. In Section \ref{sec:Objective_QA} we provide a detailed description of our proposed VQA model. In Section \ref{sec:experiments} we report and analyze various experimental results, and provide some concluding remarks in Section \ref{sec:conclusion}.

\section{Proposed Method}
\label{sec:Objective_QA}

\begin{figure}[t]
    \centering
    \includegraphics[width=0.45\textwidth]{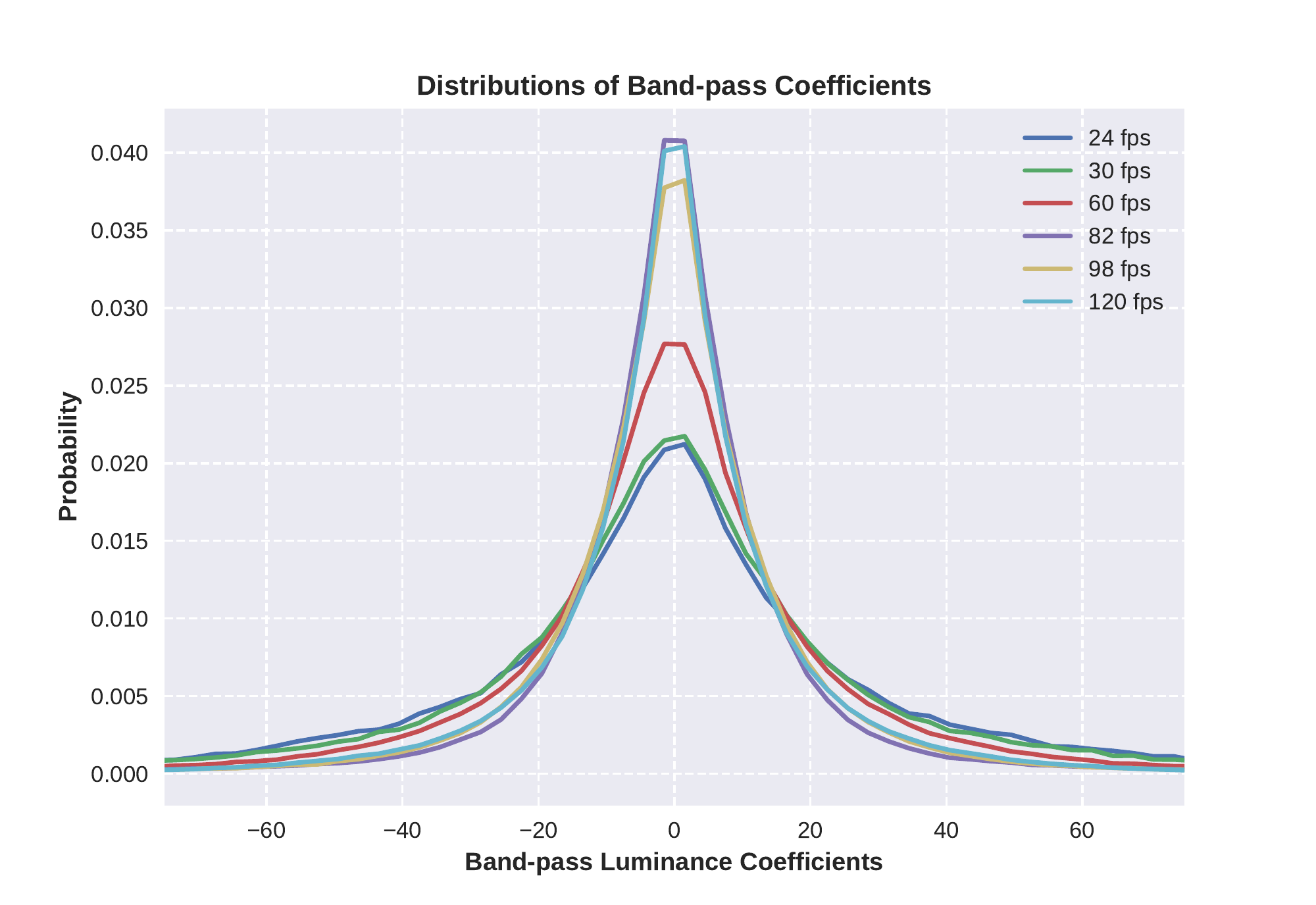}
    \caption{Distributions of band-pass coefficients for six different frame rates}
    \label{fig:fps_comparison}
\end{figure}

Consider a bank of $K$ temporal band-pass filters denoted by $b_k$ for $k \in \{1,\ldots K\}$, the temporal band-pass response for a video $V(\mathbf{x},t)$ ($\mathbf{x} = (x,y)$ represents spatial co-ordinates and $t$ denotes temporal dimension) is given by
\begin{align}
    B_k(\mathbf{x},t) = V(\mathbf{x},t)*b_k(t) \text{\hspace{10pt}} \forall k \in \{1,\ldots K\},
\end{align}
where $B_k$ denotes band-pass response of $k^{th}$ filter. Note that these are 1D filters applied only along the temporal dimension. We empirically observe that the distributions of the coefficients of $B_k$ vary as a function of frame rate. This is illustrated in Fig. \ref{fig:fps_comparison}, where distributions at different frame rates are shown for a 4-level Haar wavelet filter. From Fig. \ref{fig:fps_comparison} it maybe observed that as frame rates increase, the distribution becomes more peakier as the correlation between the consecutive frames increase with frame rate. Since coefficients of $B_k$ are band-pass in nature, they can be well modelled as following a Generalized Gaussian Distribution (GGD). GGD models have been widely used to model band-pass coefficients in many previous applications, such as image denoising \cite{chang2000adaptive}, texture retrieval \cite{do2002wavelet} etc. In this work we propose to employ entropic differences of band-pass GGD samples to quantify the deviations in distribution of band-pass coefficients. 

\vspace{-7pt}
\subsection{GGD based statistical model}
\label{subsec:GGD_model}
Let the reference and distorted videos be denoted by $R$ and $D$ respectively, with $R_t,D_t$ representing corresponding frames at time $t$. Note that $R$ and $D$ can have different frame rates though we require them to have same spatial resolution. Let the response of the $k^{th}$ band pass filter $b_k$, $k\in \{1,2,\ldots K\}$ on reference and distorted videos be denoted by $B_{kt}^R$ and $B_{kt} ^D$ respectively. We assume that every frame of $B_{kt}^R$, $B_{kt}^D$ follows a GGD model with zero-mean. We divide each frame into $P$ spatial blocks each of size $\sqrt{M} \times \sqrt{M}$. Let $B_{kpt}^R$ and $B_{kpt}^D$ denote vector of band pass coefficients in block $p$ for subband $k$ and frame $t$ for reference and distorted respectively. We allow the band-pass coefficients to pass through a Gaussian channel to model perceptual imperfections such as neural noise \cite{sheikh2006image,soundararajan2012video}. Let $\Tilde{B}_{kpt}^R,\Tilde{B}_{kpt}^D$ represent coefficients which undergo channel imperfections to obtain the observed responses $B_{kpt}^R,B_{kpt}^D$ respectively. Also let $\Tilde{B}_{kpt}^R,\Tilde{B}_{kpt}^D$ both be modeled as following GGD. This model is expressed as:
\begin{align}
    B_{kpt}^R = \Tilde{B}_{kpt}^R + W_{kpt}^R \quad B_{kpt}^D = \Tilde{B}_{kpt}^D + W_{kpt} ^D
    \label{eqn:neural_noise}
\end{align}
where $\Tilde{B}_{kpt}^R$ is independent of $W_{kpt}^R$, $\Tilde{B}_{kpt}^D$ is independent of $W_{kpt}^D$, and where $W_{kpt}^R$, $W_{kpt}^D$ are drawn from the Gaussian distribution $\mathcal{N}(0,\sigma_W ^2 \mathbf{I_M})$. It can be inferred from (\ref{eqn:neural_noise}) that $B_{kpt}^R,B_{kpt}^D$ need not necessarily be GGD, although it can be well approximated by a GGD \cite{zhao2004sum} due to the independence assumption. As shown in the VIF \cite{sheikh2006image} formulation, distortion results in a loss of "natural" image information as measured by suitably defined entropies. Variations over time of video frames from distortion can affect this visual information flow, and may depend on frame rate. For example, a lower frame rate may result in judder, which measurably affects the information flow, as measured by entropy under the statistical model of videos. The entropy of a GGD random variable $X \sim GGD(0,\alpha,\beta)$ has a closed form expression given by:
\begin{align}
    h(X) = \frac{1}{\beta} - \log \left(\frac{\beta}{2\alpha\Gamma(1/\beta) }\right)
    \label{eqn:ggd_entropy}
\end{align}
where $\alpha$ and $\beta$ are the scale and shape parameters of GGD respectively. Entropy computation requires the values of the GGD parameters of $\Tilde{B}_{kpt}^R$ and $\Tilde{B}_{kpt}^D$. However we only have access to $B_{kpt}^R$ and $B_{kpt}^D$. In order to estimate these parameters we follow the kurtosis matching procedure detailed in \cite{soury2015new} from which kurtosis values of $\Tilde{B}_{kpt}^R$and $\Tilde{B}_{kpt}^D$ can be obtained. The GGD parameters and kurtosis follow a bijective mapping \cite{soury2015new} where the kurtosis of a GGD random variable is given by:
\begin{align}
    Kurtosis(X) = \frac{\Gamma(5/\beta)\Gamma(1/\beta)}{\Gamma(3/\beta)^2} 
    \label{eqn:param_beta}
\end{align}

A simple grid search can be used to estimate the shape parameter $\beta$ from obtained kurtosis value. The other parameter $\alpha$ can be obtained using the relation 
\begin{align}
    \alpha = \sigma \sqrt{\frac{\Gamma(1/\beta)}{\Gamma(3/\beta)}}
    \label{eqn:param_alpha}
\end{align}
Plugging the parameters obtained from (\ref{eqn:param_beta}) and (\ref{eqn:param_alpha}) in (\ref{eqn:ggd_entropy}), the entropies $h(\Tilde{B}_{kpt}^R)$ and $h(\Tilde{B}_{kpt}^D)$ can be computed. In the next section we show how these entropies can be effectively used to assess the quality of videos.
\vspace{-7pt}
\subsection{Temporal Measure}
We define entropy scaling factors given by:
\begin{align*}
    \gamma_{kpt} ^R = \log(1+\sigma^2(\Tilde{B}_{kpt}^R)), \quad \gamma_{kpt}^D = \log(1+\sigma^2(\Tilde{B}_{kpt}^D))
\end{align*}
These scaling factors are similar to the ones used in \cite{soundararajan2012rred,soundararajan2012video}. Scaling factors lend a more local nature to our model and provide numerical stability on regions having low variance, where the entropy estimates are less stable. The entropies are modified by premultiplying with the scaling factors as shown in (\ref{eqn:temporal_scaled_entropy}). Regions having low variances will have small scaling factors, reducing the impact of noise on the entropy values:
\begin{align}
    \epsilon_{kpt} ^R = \gamma_{kpt} ^R h(\Tilde{B}_{kpt}^R), \quad \epsilon_{kpt} ^D = \gamma_{kpt} ^D h(\Tilde{B}_{kpt}^D).
    \label{eqn:temporal_scaled_entropy}
\end{align}
There exists a frame rate bias associated with the entropy values where different frame rates have entropies at different \textit{scales}. High frame rate sequences such as 120fps have much lower entropy values when compared to lower frame rates such as 24 fps, 30 fps etc. Thus simple entropy subtraction measures the difference between the frame rates of $R$ and $D$. Though this is desirable, this can be inefficient when comparing videos which only differ by compression artifacts. To remove this bias, we employ an additional video sequence termed Pseudo Reference ($PR$) signal, which is obtained by temporally downsampling the reference to match the frame rate of the distorted video. In our implementation we use frame dropping to conduct temporal downsampling using the FFmpeg \cite{ffmpeg} tool. In the case when the distorted sequence has the same frame rate as the reference, $PR$ will be the same as $R$. Similar to $\epsilon_{kpt}^R$ and $\epsilon_{kpt}^D$, we calculate $\epsilon_{kpt} ^{PR}$. We define the Generalized Temporal Index (GTI) as:
\begin{align}
    GTI_{kt} = \frac{1}{P}\sum_{p=1} ^P \Bigg|\Big(1 + |\epsilon_{kpt} ^D - \epsilon_{kpt} ^{PR}|\Big) \frac{\epsilon_{kpt} ^R + 1}{\epsilon_{kpt} ^{PR} + 1} - 1 \Bigg|.
    \label{eqn:GTI}
\end{align}
(\ref{eqn:GTI}) can be interpreted by decomposing into two factors: absolute difference term and ratio term. Absolute difference term removes frame rate bias and captures the quality changes as if $R$ and $D$ were at the same frame rate. The ratio term weights these factors depending on the reference and distorted frame rate. In the case of reference and distorted videos having same frame rate, the ratio term will be 1, thus making (\ref{eqn:GTI}) depend only on absolute difference. The unit terms within the absolute values ensure that $GTI$ does not become zero when $D = PR \neq R$, which happens when distorted sequence is temporally subsampled version of the reference. Note that $GTI = 0$ only when $D = PR = R$. The unit terms in the ratio avoid indeterminate values in regions having small entropy values.

\vspace{-7pt}
\subsection{Spatial Measure}
Although $GTI$ does capture spatial information due to its spatial block based nature, it is primarily influenced by the temporal filtering. To extract information about spatial artifacts, we employ spatial band-pass filters applied to every frame of the video. For this purpose we employ a local Mean Subtracted (MS) filtering similar to \cite{bampis2017speed}. Let $R_t ^{MS} = R_t - \mu_t ^R$ and $D_t ^{MS} = D_t - \mu_t ^D$ be the reference and distorted MS coefficients where local mean is calculated as
\begin{align*}
\begin{aligned}
    \mu_t ^R(i,j) = \sum_{g=-G} ^G \sum_{h=-H} ^H \omega_{g,h} R_t(i+g,j+h), \\ \mu_t ^D(i,j) = \sum_{g=-G} ^G \sum_{h=-H} ^H \omega_{g,h} D_t(i+g,j+h)
\end{aligned}
\end{align*}
where $\omega = {\omega_{g,h}|g = -G,\ldots G, h = -H,\ldots H}$ is a 2D
circularly symmetric Gaussian weighting function sampled out to 3 standard deviations. In our implementation we use $G = H = 7$. The MS coefficients $R_t ^{MS}$, $D_t ^{MS}$ are modeled as following a GGD model. Similar to the temporal measure, we divide each frame into $P$ nonoverlapping blocks and calculate entropies $h(\Tilde{R}_t ^{MS})$ and $h(\Tilde{D}_t ^{MS})$ as detailed in subsection \ref{subsec:GGD_model} by replacing temporal band-pass responses with corresponding MS coefficients. Similarly we define scaling factors and modified entropies:
\begin{align*}
\begin{aligned}
    \eta_{pt} ^R = \log(1+\sigma^2(\Tilde{R}_{pt}^{MS}))&, \quad \eta_{pt}^D = \log(1+\sigma^2(\Tilde{D}_{pt}^{MS})) \\
    \theta_{pt} ^R = \eta_{pt} ^R h(\Tilde{R}_{pt}^{MS})&, \quad \theta_{pt} ^D = \eta_{pt} ^D h(\Tilde{D}_{pt}^{MS}).
    \end{aligned}
\end{align*}
Since spatial entropies are computed using only the information from a single frame, the values are frame rate agnostic. Thus there does not arise any scale variations due to frame rate, as seen in the temporal case. The Generalized Spatial Index (GSI) is then defined as:
\begin{align}
    GSI_t = \frac{1}{P}\sum_{p=1} ^P |\theta_{pt} ^D - \theta_{pt} ^R|.
    \label{eqn:GSI}
\end{align}

\vspace{-7pt}
\subsection{Spatio-temporal Measure}
GSI and GTI operate individually on data obtained by separate processing of spatial and temporal frequency responses. Interestingly, while GSI is obtained in a purely spatial manner, GTI has both spatial and temporal information embedded in it (as entropies are obtained in a spatial blockwise manner). Thus temporal artifacts such as judder etc. only influence GTI, while spatial artifacts affect both GTI and GSI. A combined Generalized Spatio-Temporal Index (GSTI) is defined as:
\begin{align}
    GSTI_{kt} = GTI_{kt} GSI_t.
    \label{eqn:spatio_temporal_score}
\end{align}
The quality score obtained from (\ref{eqn:spatio_temporal_score}) provides scores at frame level. To obtain a video level quality score we average pool (tacitly assuming frames are temporally consistent, i.e., do not contain scene cuts, which are easily detected) the frame scores:
\begin{align}
    GSTI_k = \frac{1}{T} \sum_{t = 1} ^T GSTI_{kt}.
    \label{eqn:GSTI}
\end{align}

\begin{table}[t]
\caption{Performance comparison of FR-VQA algorithms on the HFR database. In each column the first and second best values are boldfaced and underlined, respectively}
    \label{Table:MOS_comparison}
    \centering
    \begin{tabular}{|c||c|c|c|c|}
        \hline
        ~    & SROCC $\uparrow$ & KROCC $\uparrow$ & PLCC $\uparrow$ & RMSE $\downarrow$ \\ \hline \hline
        PSNR & 0.6950 & 0.5071 & 0.6685 & 9.023 \\ 
		SSIM \cite{wang2004image} & 0.4494 & 0.3102 & 0.4526 & 10.819 \\ 
		MS-SSIM \cite{wang2003multiscale} & 0.4898 & 0.3407 & 0.4673 & 10.726 \\ 
		FSIM \cite{zhang2011fsim} & 0.4469 & 0.3151 & 0.4435 & 10.874 \\ 
		ST-RRED \cite{soundararajan2012video} & 0.5531 & 0.3800 & 0.5107 & 10.431 \\ 
		SpEED \cite{bampis2017speed} & 0.4861 & 0.3409 & 0.4449 & 10.866 \\ 
		FRQM \cite{zhang2017frame} & 0.4216 & 0.2956 & 0.452 & 10.804 \\ 
		VMAF \cite{VMAF2016}& \underline{0.7303} & \underline{0.5358} & \underline{0.7071} & \underline{8.587} \\
		deepVQA \cite{kim2018deep} & 0.3463 & 0.2371 & 0.3329 & 11.441 \\
        GSTI (Ours) & \textbf{0.7909} & \textbf{0.5979} & \textbf{0.7910} & \textbf{7.422} \\
        \hline
    \end{tabular}
\end{table}

\begin{table*}[t]
\caption{Performance comparison of various FR methods for individual frame rates in the LIVE-YT-HFR database. In each column first and second best values are boldfaced and underlined, respectively}
\label{Table:FPS_comparison}
\centering
	\scriptsize
	\scalebox{0.9}{
    \begin{tabular}{|c||c|c|c|c|c|c|c|c|c|c|c|c|c|c|}
    \hline
      & \multicolumn{2}{|c|}{24 fps} & \multicolumn{2}{|c|}{30 fps} & \multicolumn{2}{|c|}{60 fps} & \multicolumn{2}{|c|}{82 fps} & \multicolumn{2}{|c|}{98 fps} & \multicolumn{2}{|c|}{120 fps} & \multicolumn{2}{|c|}{Overall} \\
        \cline{2-15}
        ~ & SROCC$\uparrow$ & PLCC$\uparrow$ & SROCC$\uparrow$ & PLCC$\uparrow$ & SROCC$\uparrow$ & PLCC$\uparrow$ & SROCC$\uparrow$ & PLCC$\uparrow$ & SROCC$\uparrow$ & PLCC$\uparrow$ & SROCC$\uparrow$ & PLCC$\uparrow$ & SROCC$\uparrow$ & PLCC$\uparrow$\\ \hline \hline
        PSNR & \underline{0.4101} & \underline{0.3647} & \underline{0.4414} & \underline{0.4179} & \underline{0.6202} & 0.5719 & \underline{0.6878} & 0.6431 & 0.7171 & 0.6489 & 0.6019 & 0.5937 & 0.6950 & 0.6685 \\ 
			SSIM \cite{wang2004image} & 0.1277 & 0.0949 & 0.1108 & 0.0816 & 0.2123 & 0.1845 & 0.2079 & 0.2430 & 0.3876 & 0.3964 & \underline{0.7485} & 0.6726 & 0.4494 & 0.4526 \\ 
			MS-SSIM \cite{wang2003multiscale} & 0.2221 & 0.1500 & 0.1929 & 0.1112 & 0.2516 & 0.1900 & 0.2906 & 0.2549 & 0.4237 & 0.4007 & 0.6165 & 0.5843 & 0.4898 & 0.4673\\ 
			FSIM \cite{zhang2011fsim} & 0.3670 & 0.3038 & 0.3208 & 0.2638 & 0.2472 & 0.2615 & 0.3225 & 0.3055 & 0.3861 & 0.2646 & 0.3056 & 0.1178 & 0.4469 & 0.4435\\ 
			ST-RRED \cite{soundararajan2012video} & 0.1541 & 0.0369 & 0.1188 & 0.0307 & 0.5062 & 0.4457 & 0.3394 & 0.3271 & 0.4962 & 0.4556 & 0.6745 & 0.5906 & 0.5531 & 0.5107\\ 
			SpEED \cite{bampis2017speed} & 0.2591 & 0.1237 & 0.2278 & 0.0896 & 0.1824 & 0.1110 & 0.2955 & 0.2425 & 0.4118 & 0.3295 & 0.6827 & 0.6097 & 0.4861 & 0.4449\\ 
			FRQM \cite{zhang2017frame} & 0.1556 & 0.2089 & 0.0983 & 0.0854 & 0.0947 & 0.0309 & 0.0137 & 0.0035 & 0.0317 & 0.0100 & - & - & 0.4216 & 0.4520\\ 
			VMAF \cite{VMAF2016} & 0.1743 & 0.2669 & 0.2855 & 0.3740 & 0.5408 & \underline{0.6015} & 0.6820 & \underline{0.7390} & \textbf{0.8214} & \textbf{0.8128} & \textbf{0.7943} & \textbf{0.7844} & \underline{0.7303} & \underline{0.7071}\\ 
			deepVQA \cite{kim2018deep} & 0.1144 & 0.0495 & 0.1353 & 0.1059 & 0.2527 & 0.1652 & 0.1803 & 0.1515 & 0.2816 & 0.2654 & 0.6865 & 0.6209 & 0.3463 & 0.3329\\
        GSTI (Ours) & \textbf{0.4538} & \textbf{0.5935} & \textbf{0.4758} & \textbf{0.6689} & \textbf{0.6552} & \textbf{0.7566} & \textbf{0.7633} & \textbf{0.8183} & \underline{0.7844} & \underline{0.7775} & 0.7390 & \underline{0.7003} & \textbf{0.7909} & \textbf{0.7910}\\
        \hline
    \end{tabular}}
\end{table*}

\textbf{Implementation Details}. For simplicity we implemented our method only in the luminance domain. We use a 3-level Haar wavelet filter as the temporal band-pass filter $b_k$ with $k \in \{1,\ldots 7\}$ (we ignore the low pass response), where a higher $k$ value denotes a larger center frequency. We used wavelet packet (constant linear bandwidth) (WP) filter bank \cite{coifman1992entropy} as we found it to be more effective than using constant octave bandwidth filters. For entropy calculation we choose spatial blocks of size $5 \times 5$ (\ie{ }$\sqrt{M} = 5$). We choose neural noise variance $\sigma_W ^2 = 0.1$ defined in (\ref{eqn:neural_noise}). Note that similar values were employed in \cite{sheikh2006image} and \cite{soundararajan2012rred}. We observed that our algorithm is most effective when spatial resolution is downsampled 16 times along both dimensions. Similar observations were made in \cite{soundararajan2012video} and \cite{bampis2017speed} and is attributed to the motion downshifting phenomenon where, in presence of motion, human vision tends to be more sensitive to coarser scales than finer ones. Since reference and distorted sequences can have different frame rates, the reference entropy terms $\epsilon_{kpt} ^R$, $\theta_{kt} ^R$ will have a different number of frames when compared to their counterpart distorted entropy terms $\epsilon_{kpt} ^D$, $\theta_{pt} ^D$. Thus we temporally average reference entropy terms as:
\vspace{-2pt}
\begin{equation*}
\begin{aligned}
{\epsilon}_{kpt}^R &\leftarrow \frac{1}{F}\sum_{n=1}^F \epsilon_{kpt'}^R
 &\quad &\raisebox{-1.5\normalbaselineskip}[0pt][0pt]{
     where \text{$
     \begin{cases}
        F &= \frac{FPS_{\textrm{ref}}}{FPS_{\textrm{dist}}}, \\
        t' &= (t - 1)F + n
      \end{cases}$
    }} \\
{\theta}_{pt}^R &\leftarrow \frac{1}{F}\sum_{n=1}^F \theta_{pt'}^R
\end{aligned}
\end{equation*}

\section{Experiments}
\label{sec:experiments}

\textbf{Experimental Settings}.
We selected 4 FR-IQA methods: PSNR, SSIM \cite{wang2004image}, MS-SSIM \cite{wang2003multiscale} and FSIM \cite{zhang2011fsim} for comparison. Since these are image indices, they are computed on every frame and averaged across all frames to obtain the video scores. In addition to the above IQA indices, we also include 5 FR-VQA indices: ST-RRED \cite{soundararajan2012video}, SpEED \cite{bampis2017speed}, FRQM\cite{zhang2017frame}, VMAF\footnote{We use the pretrained VMAF model available at \url{https://github.com/Netflix/vmaf}} \cite{VMAF2016} and deepVQA \cite{kim2018deep}. For deepVQA, we use only stage-1 of the pretrained model (trained on the LIVE-VQA \cite{seshadrinathan2010study} database) obtained from the code released by the authors. Since the above methods require same frame rates for reference and distorted videos, for cases with differing frame rates, the distorted video was temporally upsampled by frame duplication to match the reference frame rate. Although we can downsample the reference as well, we avoided this method since it can potentially introduce artifacts (\eg{ } judder) in the reference video which is not desirable. All the above VQA models were evaluated at their original spatial resolution. Spearman's rank order correlation coefficient (SROCC), Kendall's rank order correlation coefficient (KROCC), Pearson's linear correlation coefficient (PLCC) and root mean squared error (RMSE) were the main performance criteria employed to evaluate the VQA methodologies. Before computing PLCC and RMSE, the predicted scores were passed through a four-parameter logistic non-linearity, as described in \cite{VQEG2000}.


\vspace{-7pt}
\subsection{Correlation Against Human Judgments}
The correlations between objective scores predicted by various FR models against the human judgments in the LIVE-YT-HFR database are compared in Table \ref{Table:MOS_comparison}. Our proposed method outperformed all the existing models across every evaluation criteria, as illustrated in Table \ref{Table:MOS_comparison}. The reported results for GSTI in Table \ref{Table:MOS_comparison} correspond to the first subband (\ie{ }$b_1$) of the band-pass filter, which was empirically observed to achieve highest performance when compared to other subbands. 

\vspace{-7pt}
\subsection{Performance Analysis with Individual Frame Rates}
In this experiment we subdivided the LIVE-YT-HFR database into sets which contain videos having the same frame rate, and individually analyzed the performance on them. The performance comparison is shown in Table \ref{Table:FPS_comparison}. To avoid clutter we only include SROCC and PLCC for evaluation. At high frame rates, there are naturally reduced temporal distortions, hence distortions are primarily from compression, which VMAF is (Pareto) optimized to handle. We also observed an interesting anomaly where PSNR achieved higher performance at lower frame rates when compared to other prior VQA models, which is surprising, since PSNR correlates poorly against human quality perception \cite{wang2009mean}. It is possible that frame-based models like SSIM, which accurately predict spatial distortions, have a "spatial bias" on this database. PSNR, which is merely a space-time difference signal will not have such a bias. For FRQM, correlation values are not reported for 120 fps, as it requires the compared videos to have different frame rates. It should be noted that a factor in the performance of FRQM (Table \ref{Table:FPS_comparison}) could be that it was designed on frame averaging, rather than frame dropping.



\section{Conclusion and Future Work}
\label{sec:conclusion}
We presented a simple, highly generalizable video quality evaluation method that can be employed when reference and distorted videos having different frame rates, and gauged its performance on the new LIVE-YT-HFR database. We performed a holistic evaluation of our method in terms of correlation against human perception and established that our method is superior and more robust than existing algorithms.

For band-pass filtering, a simple Haar filter was used, which can potentially limit performance. As part of future work we plan to explore other band-pass filters with superior frequency responses. Another avenue we wish to explore is to incorporate GSTI into a data driven quality model such as VMAF \cite{VMAF2016}, to further enhance performance.

\ifCLASSOPTIONcaptionsoff
  \newpage
\fi

\bibliographystyle{IEEEtran}
\bibliography{template_4}

\end{document}